\documentclass[review]{elsarticle}

\usepackage{lineno,hyperref}

\usepackage{amsmath}
\usepackage{amsthm}
\usepackage{amssymb}
\usepackage{amsfonts}

\usepackage{soul}
\usepackage{graphicx}% Include figure files
\usepackage{dcolumn}% Align table columns on decimal point
\usepackage{bm}% bold math
\usepackage[usenames,dvipsnames]{color}
\usepackage[normalem]{ulem}
\modulolinenumbers[5]

\DeclareMathOperator{\arctanh}{arctanh}

\newcommand{\be}{\begin{equation}}
\newcommand{\ee}{\end{equation}}
\newcommand{\bq}{\begin{eqnarray}}
\newcommand{\eq}{\end{eqnarray}}

\newcommand{\bfx}{\mathbf{x}}
\newcommand{\bfy}{\mathbf{y}}

\journal{Annals of Physics}
\begin{document}
\begin{frontmatter}

\title{On the dispute between Boltzmann and Gibbs entropy}% Force line breaks with \\
%\thanks{A footnote to the article title}%

\author{Pierfrancesco Buonsante}
%\affiliation{
% QSTAR \& CNR - Istituto Nazionale di Ottica, 
% Largo Enrico Fermi 2, I-50125 Firenze, Italy. 
%}
% \altaffiliation[Also at ]{QSTAR \& CNR - Istituto Nazionale di Ottica, 
% Largo Enrico Fermi 2, I-50125 Firenze, Italy.}%Lines break automatically or can be forced with \\
\author{Roberto Franzosi}%
%\affiliation{
% QSTAR \& CNR - Istituto Nazionale di Ottica, 
% Largo Enrico Fermi 2, I-50125 Firenze, Italy. 
%}
\ead{roberto.franzosi@ino.it}
\author{Augusto Smerzi}
\address{
 QSTAR \& CNR - Istituto Nazionale di Ottica, 
 Largo Enrico Fermi 2, I-50125 Firenze, Italy. 
}

%\collaboration{MUSO Collaboration}%\noaffiliation

%\author{Charlie Author}
% \homepage{http://www.Second.institution.edu/~Charlie.Author}
%\affiliation{
% Second institution and/or address\\
% This line break forced% with \\
%}%
%\affiliation{
% Third institution, the second for Charlie Author
%}%
%\author{Delta Author}
%\affiliation{%
% Authors' institution and/or address\\
% This line break forced with \textbackslash\textbackslash
%}%

%\collaboration{CLEO Collaboration}%\noaffiliation

\date{\today}% It is always \today, todayt,
             %  but any date may be explicitly specified

\begin{abstract}

The validity of the concept of negative temperature has been recently challenged by arguing that
the Boltzmann entropy 
(that allows negative temperatures) is inconsistent from a mathematical and statistical
point of view, whereas the Gibbs entropy (that does not admit negative temperatures)
provides the correct definition for the microcanonical entropy.
Here we prove that 
the Boltzmann entropy is thermodynamically and mathematically consistent.
Analytical results on two systems supporting negative temperatures illustrate the scenario
we propose.
In addition we numerically study a lattice system to show that negative temperature equilibrium
states are accessible and obey standard statistical mechanics prediction.

\begin{description}
%\item[Usage]
%Secondary publications and information retrieval purposes.
\item[PACS numbers] 05.20.-y, 05.20.Gg, 05.30.-d, 05.30.Ch
%May be entered using the \verb+\pacs{#1}+ command.
%\item[Structure]
%You may use the \texttt{description} environment to structure your abstract;
%use the optional argument of the \verb+\item+ command to give the category of each item. 
\end{description}
\end{abstract}

\begin{keyword}
Statistical Mechanics, Microcanonical Ensemble
%\texttt{elsarticle.cls}\sep \LaTeX\sep Elsevier \sep template
%\MSC[2010] 00-01\sep  99-00
\end{keyword}

\end{frontmatter}

%\linenumbers

%\pacs{05.20.Gg, 02.40.Vh, 05.20.- y, 05.70.- a}% PACS, the Physics and Astronomy
                             % Classification Scheme.
%\keywords{Statistical Mechanics}%Use showkeys class option if keyword
                              %display desired
%\maketitle

%\tableofcontents

\section{\label{secIntro}Introduction}
The concept of negative absolute temperature  
was\footnote{This was not the first place where negative
temperatures have been considered, in fact \cite{Onsager49} two years before proposed the existence
of negative temperatures in order to explain the formation of large scale vortices by clustering of small
ones in hydrodynamic systems.} invoked to explain the results of experiments  with nuclear-spin
systems carried out by Pound \cite{Pound}, Purcell and Pound \cite{PurcellPound} and
Ramsey and Pound \cite{RamseyPound}. Shortly after these experiments, Ramsey
\cite{Ramsey} discussed the thermodynamic implications of  negative absolute temperature and their
meaning in statistical
mechanics, thereby granting this concept a well-grounded place in physics~\cite{Kittel,Landau}. 

The microcanonical ensemble, which provides the statistical description of an isolated system at equilibrium,
is the most appropriate venue to discuss negative temperatures. 
In this ensemble, the thermodynamic quantities, like temperature and specific heat,
are derived from the entropy through suitable thermodynamic relations. For instance, the inverse temperature is
proportional to the derivative of the entropy with respect to the energy.  
In equilibrium statistical mechanics, there are at least two commonly accepted definitions of entropy:
the Boltzmann entropy is proportional to the logarithm of the number of microstates
in a given ``energy shell'', whereas the Gibbs entropy is
proportional to the logarithm of the number of
microstates up to a given energy.
The debate as to which of these definitions of entropy is the correct one has been going on for a long time
\cite{Jaynes,Hertz10,Einstein11,Schluter48,Berdichevsky91,Munster_1987,Pearson85,
Campisi05,Adib04,Lavis2005245},
although the general consensus is that they are basically interchangeable. In fact, for
standard systems\footnote{ 
With ``standard system'' we mean a system with unbounded energy from above for which
the energy goes to infinity when one of the canonical coordinates goes to infinity.}
with a large number of degrees of freedom they are practically equivalent \cite{Huang_1987}.
Full equivalence is obtained in  the thermodynamic limit.

The existence of negative-temperature
states provides a major bone of contention in the debate. In fact, negative temperatures emerge
in the Boltzmann description whenever the number of microstates in a given energy shell is a
decreasing function of the relevant energy. On the contrary, the Gibbs temperature can never
be negative, since the number of microstates having energy below a given value is always a
non-decreasing function of such value. Thus, systems admitting negative (Boltzmann)
temperatures provide an ideal context to address the matter of the correct definition
of entropy.

Recently \cite{Dunkel2013,Hilbert_PRE_2014,Hanggi_15}  it was argued that
for a broad class of physical systems, including standard classical Hamiltonian systems,
only the Gibbs entropy yields a consistent thermodynamics, and that, consequently, negative temperatures
are not achievable within a standard thermodynamical framework.
In this respect, what is usually referred to Boltzmann {\it temperature} would not
possess the required properties for a temperature
\cite{Dunkel2013,Sokolov_2014,DunkelHilbertRep1,DunkelHilbertRep2,Hilbert_PRE_2014,Campisi_2015,Campisi_2016}.
These and other related arguments \cite{Romero-Rochin,Treumann_2014,Treumann_2014a} have been
contended \cite{Vilar_2014,Frenkel_2015,Schneider_2014,Wang_2015,Swendsen_Wang_Physicaa_2016},
in what has become a lively debate.
 
In the present manuscript, at first we focus on the class of systems whose canonical
(or, possibly, grand canonical) ensemble is equivalent to the microcanonical ensemble.
Thus we implicitly exclude non-extensive systems, and systems at the first-order phase
transitions.
We show that such equivalence can be rigorously satisfied only if the thermodynamics
of the latter ensemble is derived by the Boltzmann entropy. For such systems we show
that the Boltzmann temperature provides
a consistent description with those of the canonical 
and grand canonical ensembles. Therefore we conclude that, also in the case of
isolated systems for which a comparison between different statistical ensembles is not
possible, the Boltzmann entropy provides the correct description.
%On the contrary, the temperature predicted by Gibbs is irreconcilable
%with that one of the latter ensembles.
\\
Later, we focus on a general system  and we  prove that the Boltzmann entropy is
thermostatistically consistent and does not violate any fundamental condition for
the microcanonical entropy.

The outline of the paper is the following.
In Sec. \ref{sec:mc} we summarize the essential features of systems in which negative
Boltzmann temperatures are expected.
In Sec. \ref{sec:dyntherm} we consider an isolated Hamiltonian system and, 
under the hypothesis of ergodicity, we show that only for the Boltzmann
entropy all the thermodynamic quantities can be measured as time averages (along the dynamics)
of suitable functions. 
In Sec. \ref{sec:Compar} we prove that, for systems whose canonical and
microcanonical ensembles
are equivalent, the thermodynamically consistent 
%correspondence between $\beta$ and $\epsilon$
definition for the temperature
is the one derived with the Boltzmann entropy.
Furthermore we show that, in the thermodynamic
limit, the Gibbs and Boltzmann temperatures do coincide when the latter is positive whereas
the inverse Gibbs temperature is identically null in the region where Boltzmann provides
negative values for the temperature.
In Sec. \ref{sec:consistency} we recall the critique of consistency of Boltzmann entropy
raised recently in literature and, in Sec. \ref{sec:proof} we prove that the Boltzmann entropy 
is consistent from a mathematical and thermodynamical point of view.
In section \ref{sec:evidence}, we give two examples of systems supporting negative
Boltzmann temperatures for which the grand-canonical (or canonical) ensemble
and the microcanonical description given by the Boltzmann entropy
do agree on the whole parameter
space and on the complete range of values of the energy-density.
We show that the equipartition theorem fails for system with negative Boltzmann temperatures in
Sec. \ref{sec:equipar}. In Sec. \ref{sec:measuring} we show through
numerical simulations on a specific system, that negative temperatures are accessible. We show that,
irrespective of the sign of the temperature, a large microcanonical lattice acts as a thermostat for
a small grand canonical sublattice, and this confirms the ensemble equivalence. Furthermore, we
have shown that, irrespective of the sign of the temperature, two isolated systems at equilibrium
at different inverse temperatures, reach an equilibrium state at an intermediate inverse temperature,
after that they are brought in contact.

\section{\label{sec:mc} Negative temperatures}
The microcanonical ensemble describes the equilibrium properties of an
isolated system, that is to say in which energy, and possibly further quantities, are conserved.
Within the microcanonical description, all the thermodynamic quantities are derived from the entropy,
for instance the inverse temperature of the system is defined as
\begin{equation}
\beta =\dfrac{1}{k_B} \dfrac{\partial s}{\partial \epsilon} \, ,
\label{betag}
\end{equation}
where $k_B$ is the Boltzmann constant and $s(\epsilon)$ is the entropy density
corresponding to a given energy density $\epsilon$. 
The two alternative definitions for the entropy used in equilibrium statistical mechanics
are ascribed to Boltzmann and Gibbs\footnote{We refer
to Ref. \cite{Muller_2009} for historical details.}. According to Boltzmann's definition
\begin{equation}
 s_B(\epsilon) = L^{-1} k_B \ln (\omega(\epsilon) \Delta) \, ,  
\label{entropyboltz}
\end{equation}
where $\omega(\epsilon)$ is the density of microstates at a fixed value energy density 
$\epsilon$ and, possibly, at a fixed value of the additional conserved quantities, $\Delta$ is a
constant with the same dimension as
$\epsilon$, and $L$ is the number of degrees of freedom in the system.
The Gibbs entropy is
\begin{equation}
 s_G(\epsilon) = L^{-1} k_B \ln \Omega  (\epsilon)\, ,  
\label{entropygibbs}
\end{equation}
where $\Omega  (\epsilon)$ is the number of microstates with energy density less than or equal to
$\epsilon$ and, possibly, at a fixed value of the additional conserved quantities. 
It is known that in the thermodynamic limit these two definitions of entropy lead to
equivalent thermodynamic results in ``standard'' systems \cite{TodaKuboSaito}.
So far, these two entropy definitions have been used
in an alternative (interchangeable) way in statistical mechanics, by resorting to the
most suitable form depending on the specific problem considered.
These two entropy definitions are connected by the relation between
$\omega$ and $\Omega$
\begin{equation}
\omega(\epsilon) = \dfrac{\partial \Omega}{\partial \epsilon}(\epsilon)  \, ,
\label{omegaOmega}
\end{equation} 
Since ${\partial \Omega}/{\partial \epsilon} \geq 0$, Gibbs' temperatures are not negative
and consequently the two entropies have incompatible outcomes if applied to systems that admit
negative Boltzmann temperatures. 

A necessary (although not sufficient) condition for a system for having negative
temperatures is the boundedness of the energy (as in the case of nuclear-spin systems discussed
by Pound \emph{et al}),
in this case a local maximum inside the system's density energy interval
of the Boltzmann entropy $s_B(\epsilon)$ is not forbidden and, both positive and
negative Boltzmann temperatures are possible.

Hamiltonians with bounded energies can also be characterized by the existence of more
than one first integral of motion and, for this reason, in addition to the statistical
mechanics of systems with one first integral, we will consider also the case of systems
with more then one first integrals. Within the latter class 
for instance there are models usually employed for describing ultracold atoms.
The possibility of observing negative
temperature states in ultracold systems, has been theoretically predicted by some authors
with different approaches \cite{Mosk_2005,Rapp_2010,Iubini_2013}
and, the experimental evidence of the existence of states for
motional degrees of freedom of a bosonic gas at negative (Boltzmann)
temperatures, have been achieved a few years ago by Braun {\it et al.} \cite{Braun2013}.
The interpretation of such experimental results has been contested in
\cite{Dunkel2013}. Successively \cite{Hilbert_PRE_2014,Hanggi_15} 
it has been argued that for a broad class of systems --that includes
all ``standard classical Hamiltonian systems''--
only the Gibbs entropy satisfies all three thermodynamic laws exactly.
These papers have engendered a glowing debate
between supporters of the Gibbs entropy
\cite{DunkelHilbertRep1,DunkelHilbertRep2,Campisi_2015,Campisi_2016,Sokolov_2014,Hilbert_PRE_2014,Treumann_2014,
Treumann_2014a,Hanggi_15,Swendsen_Wang_Physicaa_2016} 
and those considering correct the Boltzmann entropy
\cite{Frenkel_2015,Schneider_2014,Vilar_2014,Poulter_2015, Cerino_2015,Buonsante_2015,
Swendsen_Wang_PRE_2015,Swendsen_PRE_2015,Swendsen_Wang_Physicaa_2016,MLGS_2015}.

\section{\label{sec:dyntherm} Dynamics and statistical mechanics for classical systems}
Let us consider first a generic classical many-particle system described by an autonomous
Hamiltonian $H (x_1 , . . . , x_{L} )$, in which the energy is the sole first integral of motion.
The Boltzmann entropy density $s_B(\epsilon)$ in this case is given by
\begin{equation}
 s_B(\epsilon) = L^{-1} k_B \ln\! \int \! d^{L}\mathbf{x} \,
\delta(L \epsilon - {H}(\mathbf{x})) \, ,  
\label{entropyb}
\end{equation}
whereas the one of Gibbs is
\begin{equation}
 s_G(\epsilon) = L^{-1} k_B \ln\! \int \! d^{L}\mathbf{x} \,
\Theta(L \epsilon - {H}(\mathbf{x})) \, ,  
\label{entropyg}
\end{equation}
where $\delta$ is the Dirac function and $\Theta$ is the Heaviside function.

As a consequence of the conservation of energy, the system dynamics takes place on energy-level sets.
From Liouville theorem it descends that the measure of the Euclidean volume is
preserved by the dynamics and this induces a measure $\mu$ conserved on each energy
level set $\Sigma_\epsilon$ of energy density $\epsilon$ which is given by
\cite{khinchin_1949,Berdichevsky_1988}
\begin{equation}
{ d \mu} = \dfrac{d \Sigma}{\Vert \nabla H \Vert} \, ,
\label{hs-measure}
\end{equation}
where $d \Sigma$ is the Euclidean measure induced on $\Sigma_\epsilon$ and $\Vert \cdot \Vert $ is
the Euclidean norm.

This means that, under the hypothesis of ergodicity, the averages of each dynamical
observable $\Phi$ of the system can be equivalently measured along the dynamics as
\begin{equation}
\langle \Phi \rangle = \lim_{\tau \to\infty} \dfrac{1}{\tau} \int_0^\tau dt \, \Phi(t) \, ,
\label{phit}
\end{equation}
or as average on the hypersurface $\Sigma_\epsilon$ according to
\begin{equation}
\langle \Phi \rangle = \dfrac{\int_{\Sigma_\epsilon} d \mu \, \Phi }{\int_{\Sigma_\epsilon} d \mu } \, .
\label{phimu}
\end{equation}
Now, it is reasonable to expect that the temperature, the specific heat, and the other
thermodynamic observables could be measured as time averages of suitable observables $\Phi$
along the dynamics, in a way analogous to Eq. (\ref{phit}).
Consequently, when ergodicity holds, the measures of these quantities have to be derived
from averages upon the energy level sets $\Sigma_\epsilon$, according to Eq. (\ref{phimu}).
Furthermore, temperature, specific heat and other thermodynamic quantities depend
on derivatives of the microcanonical entropy with respect to energy. Therefore, they are computed
by means of a functional of the form (\ref{phimu}) \emph{if and only if} the microcanonical
entropy is defined \`a la Boltzmann.
This fact is proven by Rugh \cite{RughPRL97} in the case of many-particle systems for which
the Hamiltonian is the only conserved quantity, and in Ref. \cite{Franzosi_JSP11} and Ref.
\cite{Franzosi_PRE12} for the case of two and $k\in \mathbb{N}$ conserved quantities,
respectively.

For instance, in the simpler case studied in Ref. \cite{RughPRL97} e.g., it results
\[
s_B = L^{-1}k_B \ln \int_{\Sigma_\epsilon} d \mu \, ,
\]
and from the definition \eqref{betag} in the case of Boltzmann we obtain
\footnote{The Federer-Laurence derivation formula 
\cite{Federer_1969,Laurence_1989} is
$
\partial^k (\int_{\Sigma_\epsilon}  \psi
 d \Sigma )/\partial \epsilon^k  = L^k\int_{\Sigma_\epsilon} A^k\left( \psi\right)
 d \Sigma
$,
where $A(\bullet) = 1~/~\|~\bigtriangledown~H~\|~\bigtriangledown
 \left( \bigtriangledown H/\|\bigtriangledown H \| \bullet \right)$.
}
\begin{equation}
\beta_B = \dfrac{\int_{\Sigma_\epsilon} d \mu \, \nabla \cdot (\nabla H/\Vert \nabla H \Vert^2) }
{\int_{\Sigma_\epsilon} d \mu } \, ,
\label{betarugh}
\end{equation}
where $\beta_B = 1/(k_B T_B)$\footnote{ Higher derivatives of $s_B$ respect to $\epsilon$
are computed by means the Federer-Laurence formula \cite{Federer_1969,Laurence_1989} that leads to
averages similar to the one in Eq. \eqref{phimu}.}.
In the case of $s_G$ the expression for the inverse temperature
is
\begin{equation}
\beta_G = \dfrac{\int_{\Sigma_\epsilon} d \mu}
{\int_{M_\epsilon} d^{L}\mathbf{x} } \, ,
\label{betagibbs}
\end{equation}
where $M_\epsilon = \{\mathbf{x} \in \mathbb{R}^L | H(\mathbf{x}) \leq L \epsilon  \}$, \emph{that cannot be
expressed in the form (\ref{phimu})}.
In other words, by adopting the Gibbs entropy definition when ergodicity holds true,
\emph{one has to trust in the very singular fact that time averages of
thermodynamic quantities
taken along the dynamics (and then on the energy level set $\Sigma_\epsilon$) coincide with the
averages of quantities taken on a set that includes all the energy levels below
to the one on which the dynamics takes place, analogously to Eq. \eqref{betagibbs} of the
inverse temperature}\footnote{In addition to the case of the inverse temperature, the same scenario
hold for the chemical potential.}. 

In the case of $\beta_G$,
one could suggest that the Gibbs temperature can
be measured as a
microcanonical average by resorting to the equipartition theorem. Nevertheless, as we will
show in Sec. \ref{sec:equipar}, for instance in the case of systems with negative Boltzmann temperatures,
the ``standard'' equipartition theorem fails. This is a first signal
of inconsistency for the Gibbs entropy.

It is worth emphasizing that in the case of $k>1$ conserved quantities, a
geometric structure similar to the one of Eq.~(\ref{betarugh}) keeps on to be valid.
In fact, in Ref. \cite{Franzosi_JSP11} it has been considered the case $k=2$ by studying
a general classical autonomous many-body Hamiltonian system, whose coordinates and
canonical momenta are indicated with $\mathbf{x}\in \mathbb{R}^L$, and for which
$V (\mathbf{x})$ is a further conserved quantity in involution with $H$. 
For such a system, the motion takes place on the manifolds ${\cal M} = \Sigma_\epsilon \cap  V_u$,
where $V_u = \{(\mathbf{x})\in \mathbb{R}^L | V(\mathbf{x}) = L u\}$ are subsets of
$\mathbb{R}^L$ where $V$ is constant. 
In Ref. \cite{Franzosi_JSP11} it is shown that
\begin{align}
s_B = &  L^{-1}k_B \ln\! \int \! d^{L}\mathbf{x} \,
\delta({H}(\mathbf{x}) -L\epsilon) \delta({V}(\mathbf{x}) -L u) 
= \nonumber \\ & 
L^{-1} k_B \ln \int_{\cal M} \dfrac{d \tau}{W} \, ,
\label{S2}
\end{align}
where $d\tau$ is the volume form of ${\cal M}$, and 
\begin{equation}
W =
\left[
\sum^L_{\mu<\nu =1}
\left(
\dfrac{\partial H}{\partial x^\mu} \dfrac{\partial V}{\partial x^\nu} -
\dfrac{\partial H}{\partial x^\nu} \dfrac{\partial V}{\partial x^\mu}
\right)^2
\right]^{1/2} \, .
\label{W}
\end{equation}
Furthermore, in \cite{Franzosi_JSP11} it is derived the generalization of (\ref{betarugh})
that gives the microcanonical inverse temperature for these systems, it results
\begin{equation}
\beta_B = \dfrac{\int_{\cal M} d \tau \, \Phi_2 }
{\int_{\cal M} d \tau } \, ,
\label{betafranz}
\end{equation}
where the complicated functional is now
\begin{equation}
\Phi_2(x) \! = \! 
		\dfrac{W}{ \nabla H \cdot n^{_\xi} } 
		\left[ 
		 \nabla \left( 
\frac{n^{_\xi}}{W} \right)
- \frac{(n^{_V} \cdot \nabla) \left( n^{_\xi} \right) }{ W} 
\cdot
n^{_V}
		\right] \, ,
\label{niceformula2}
\end{equation}
that is given in terms of the unitary vectors $n^{_H}={\nabla H}/{\Vert \nabla H \Vert }$
and $n^{_V}={\nabla V}/{\Vert \nabla V \Vert }$
through the vector
$
 \xi = n^{_H} - ( n^{_H} \cdot n^{_V}  ) n^{_V} 
$, from which is defined the unitary vector $n^{_\xi} = \xi/\Vert \xi \Vert$
that appears in Eq. (\ref{niceformula2}).
\emph{Remarkably, by exchanging $H$ and $V$ in expression \eqref{niceformula2} the functional so
obtained allows to measure the chemical potential of the system.
This fact shows a further ``esthetic advantage'' of the Boltzmann entropy: it leads to
expressions formally identical independently from the number of conserved quantities.
}

\section{\label{sec:Compar} Comparison between statistical ensembles}
In a statistical description of a many-body system, temperature has a different
meaning depending on the statistical ensemble. In the canonical ensemble and in the
grand-canonical one, (inverse) temperature is just a Lagrangian parameter that is introduced
in order to fix the mean energy.
On the contrary, in the microcanonical ensemble the temperature is a quantity derived
from the entropy density $s$, according to the relation $T = (\partial s / \partial \epsilon)^{-1}$.
Therefore it is clear that $T(\epsilon)$ will depend on the entropy definition
assumed within the microcanonical statistical description.
The main point here is that the meaning of temperature cannot be reduced to the issue
of the coherent definition inside to microcanonical ensemble, at least if there is equivalence
of ensembles. In the latter case, one expects that temperature, or more in general thermodynamics, defined
for a system by two microscopic models, for instance canonical and microcanonical, coincide in the thermodynamic
limit and they coincide with the experimentally known thermodynamics of such system \cite{Gallavotti}.
This amounts to requiring that the thermodynamics of a large isolated (microcanonical) system and the thermodynamics
of a ``small'' (even if big enough) subsystem of it coincide. In fact, in the thermodynamic limit, 
the complement of the subsystem, acts on it as a thermostat and the subsystem is well described in the canonical ensemble.
The problem of equivalence of ensembles is only incompletely solved \cite{Ruelle}, for instance it is known
that systems with long-range interaction can violate this equivalence. In fact, for this class
of systems the energy is not extensive: a system cannot be divided into independent macroscopic parts
at variance of the case of the short-range interaction.
In the following we show that if there is equivalence between statistical ensembles,
Helmholtz free energy density is the Legendre transform of Boltzmann entropy density and vice versa.
Consequently, thermodynamics derived for a systems by Boltzmann entropy and by canonical partition
function  rigorously coincide in the thermodynamic limit.
We consider this as a strong evidence supporting the legitimacy of the Boltzmann
entropy.
Let us now discuss about a case where there is not equivalence between canonical and microcanonical ensembles.
This is the case of a system with long-range interaction that undergoes a first-order phase transition.
We refer to \cite{Huller_1994} for details. In summary the Boltzmann entropy for a system with
these features is not a concave function,  consequently it cannot be the Legendre transformation  of the
Helmholtz free-energy density, and $\beta_B(\epsilon)$ is a not-invertible function.
In cases like this, the canonical ensemble has not foundation since it cannot be derived from the
microcanonical ensemble, unlike the case of the extensive systems \cite{Gross_2001}, where it can.
Therefore, the case of the long-range interactions are outside the class of systems to which our
proof applies, although we consider Boltzmann entropy the correct definition also for this
class of systems.

In the following we will consider two explicit systems, one of which has two conserved
quantities, accordingly in this section we give our proof for a system with this feature.
The restriction of our derivation to the case of a system where energy is the sole conserved
quantity is straightforward.
Let us consider an arbitrary classical many-body Hamiltonian system with $k=2$ first
integrals of motion, $H$ and a further conserved quantity $V$ which is in involution with
$H$.
In order to compare the canonical and the microcanonical description for such
class of systems, we decompose the canonical partition function as follows
\cite{Gallavotti}
\begin{align}
Z (\beta,L) =& \int d^L {\bf x} e^{-\beta H({\bf x})} \delta(L u-V({\bf x}))  = \nonumber
\\
%&L \int d\epsilon d^L {\bf x} e^{-\beta H({\bf x})} \delta(L \epsilon-H({\bf x})) \delta(L u-V({\bf x}))  =
%\nonumber \\ &
%L \int d\epsilon d^L {\bf x} e^{-\beta L \epsilon} \delta(L \epsilon -H({\bf x})) \delta(L u-V({\bf x})) = \nonumber
% \\ \nonumber
%& 
%%\int^{E_M}_{E_m} d E
%%e^{-\beta E} 
%%\int_{\cal M} \dfrac{d \tau}{W}= \nonumber
%L \int^{\epsilon_M}_{\epsilon_m} d \epsilon
%e^{-\beta L \epsilon} 
%\int_{\cal M} \dfrac{d \tau}{W} = \nonumber \\
%&= L \int^{\epsilon_M}_{\epsilon_m} d \epsilon
%e^{-\beta L \epsilon} e^{\ln (\int_{\cal M} \dfrac{d \tau}{W})} = 
%\nonumber \\ 
&=
L
 \int^{\epsilon_M}_{\epsilon_m} d \epsilon e^{-\beta L \left(\epsilon -
 \frac{s_{^{_{\scriptscriptstyle B}}}(\epsilon)}{k_B \beta} \right)}
% \approx 
% \nonumber \\ & 
%L \sqrt{\dfrac{2\pi k_B}{- L s^{\prime \prime}_{^{_{\scriptscriptstyle B}}} }}
%e^{-\beta L \left(\epsilon^* -
% \frac{s_{^{_{\scriptscriptstyle B}}}(\epsilon^*)}{k_B\beta} \right)}\, ,
\label{c2mc2}
\end{align}
where $\epsilon_m$ and $\epsilon_M$ are the minimum and the maximum of the admitted energy density
$\epsilon :=E/L$, respectively and \emph{$s_{^{_{\scriptscriptstyle B}}}(\epsilon)$ is exactly
Boltzmann's microcanonical
entropy density}. Furthermore, note that we have made use of the generalization \cite{Franzosi_JSP11}
of the co-area formula \cite{Federer_1969} which is of very general validity and holds also
for Hausdorff measurable sets.
It is worth emphasizing that in Eq. (\ref{c2mc2}) $\beta$ represents just a (Lagrangian)
parameter and it is only thanks to the comparison between canonical and
microcanonical ensemble that one can ascribe to $\beta$ the meaning of inverse
temperature \cite{Gallavotti}.
In order to connect the canonical description to the microcanonical description one has to observe
that, roughly speaking, the partition function $Z_L$ depends on the competition
between the two terms $e^{-\beta \epsilon L}$ and $e^{L {s_{^{_{\scriptscriptstyle B}}}}/{k_B}} $ which
are exponentially decreasing and increasing with $L$, respectively.
Thus, by the saddle point/Laplace method, the following asymptotic
approximation ($L \gg 1$) for the partition function holds
\begin{equation}
Z_L(\beta) \approx L \sqrt{\dfrac{2\pi k_B}{- L s^{\prime \prime}_{B}} }
e^{-\beta L \left(\epsilon^* -
 \frac{s_{^{_{\scriptscriptstyle B}}}(\epsilon^*)}{k_B\beta} \right)}
\label{c3mc3}
\end{equation}
where
$s^{\prime \prime}_{^{_{\scriptscriptstyle B}}}  = \frac{\partial^2 s_{^{_{\scriptscriptstyle B}}}}{\partial
\epsilon^2}(\epsilon^*)$, and $\epsilon^*:=\epsilon(\beta)$
is the solution of 
\begin{equation}
\beta = \dfrac{1}{k_B} \dfrac{\partial s_{^{_{\scriptscriptstyle B}}}}{\partial \epsilon}(\epsilon) \, .
\label{beta-epsilon}
\end{equation}
Therefore the canonical free energy $f$ is
\begin{equation}
f(\beta) := \lim_{L\to \infty} -\frac{1}{\beta L} \ln Z_L (\beta) = { \left(\epsilon^*
 -\dfrac{s^{^{_\infty}}_{^{_{\scriptscriptstyle B}}}(\epsilon^*)}{k_B\beta} \right)} \, ,
\label{ZTdL}
\end{equation}
where $\beta$ and $\epsilon^*$ are related by
Eq. (\ref{beta-epsilon}), \emph{the
Boltzmann definition of microcanonical temperature}.
In other words, the thermodynamic limit of the dimensionless Boltzmann
entropy $s^{^{_\infty}}_{^{_{\scriptscriptstyle B}}}(\epsilon)/k_B$, as a function of the density
energy $\epsilon$, and the
dimensionless Helmholtz free energy $\beta f(\beta)$, as a function
of the inverse Boltzmann absolute
temperature $\beta$, are connected by a Legendre transformation
\begin{equation}
\beta f(\beta) 
= \inf_{\epsilon} { \left(\beta \epsilon
 -s^{^{_\infty}}_{^{_{\scriptscriptstyle B}}}(\epsilon)/k_B \right)} \, ,
\label{Legendre}
\end{equation}
and this relation is valid only in the case of Boltzmann's definitions.
This fact shows that whenever there is equivalence between the canonical and the
microcanonical ensemble, \emph{the only consistent definition for the microcanonical temperature
is the Boltzmann's}.

It is worth remarking a general scenario in which negative temperatures emerge.
From Eqs. \eqref{omegaOmega} and \eqref{entropyb} it follows
\begin{equation}
\Omega (\epsilon) = \int^\epsilon_{\epsilon_{min}} d\epsilon^\prime
e^{L s_B(\epsilon^\prime)/k_B}  \, ,
\end{equation}
thus when $s_B(\epsilon)$ has a local maximum at $\tilde{\epsilon}$ by using
the Laplace method we deduce the following asymptotic
approximate ($L \gg 1$) expressions
\begin{eqnarray}
\Omega(\epsilon) 
\approx \dfrac{k_B}{L s^\prime_B(\epsilon)} e^{L s_B(\epsilon)/k_B} \, , \quad \epsilon < \tilde{\epsilon}
\label{eq:Omega1}\\
\Omega(\epsilon) 
\approx \sqrt{-\dfrac{2 \pi k_B}{L s^{\prime\prime}_B(\tilde{\epsilon})}} e^{L s_B(\tilde{\epsilon})/k_B} \, , 
\quad \epsilon > \tilde{\epsilon} \, ,
\label{eq:Omega2}
\end{eqnarray}
which, in the thermodynamic limit, yields
\begin{eqnarray}
\beta_G(\epsilon) := \beta_B(\epsilon) \, , \quad \epsilon < \tilde{\epsilon} 
\label{eq:OmegaLT1}
\\
\beta_G(\epsilon) := 0 \, , \quad \epsilon > \tilde{\epsilon} \, .
\label{eq:OmegaLT2}
\end{eqnarray}
\emph{The peculiar behaviour just here summarized shows in which way the Gibbs entropy and the
Boltzmann entropy are inequivalent in the thermodynamic limit in the case of systems that
allow negative Boltzmann temperatures}.

\section{ \label{sec:consistency} Critique of consistency of Boltzmann entropy}
In the present section, we focus on what we consider the heart of the
matter about the issue of the correct microcanonical entropy definition. 
In Ref. \cite{Dunkel2013} the following equations are reported:
\begin{equation}
\left(\dfrac{\partial s}{\partial \epsilon} \right)_{\bf a}^{-1} 
\left(\dfrac{\partial s}{\partial a_\mu} \right)_{\epsilon,a_\nu \neq a_\mu}
\!\!\!\!\!\!\!\!\!\!\!\!\!\!\!
 = -
\left(\dfrac{\partial \epsilon }{\partial a_\mu} \right)_{s,a_\nu \neq a_\mu}
\!\!\!\!\!\!\!\!\!\!\!\!\!\!\!
% = - 
%\left\langle \dfrac{\partial h }{\partial a_\mu}\right\rangle \, ,
\label{seia}
\end{equation}
\begin{equation}
\left(\dfrac{\partial s}{\partial \epsilon} \right)_{\bf a}^{-1} 
\left(\dfrac{\partial s}{\partial a_\mu} \right)_{\epsilon,a_\nu \neq a_\mu}
\!\!\!\!\!\!\!\!\!\!\!\!\!\!\!
% = -
%\left(\dfrac{\partial \epsilon }{\partial a_\mu} \right)_{s,a_\nu \neq a_\mu}
%\!\!\!\!\!\!\!\!\!\!\!\!\!\!\!
 = - 
\left\langle \dfrac{\partial h }{\partial a_\mu}\right\rangle \, ,
\label{seib}
\end{equation}
that are therein considered fundamental thermostatistical self-consistency conditions.
Here $a_\mu$ are intensive parameters of the Hamiltonian density
$h=H/L$  and $\langle \cdot \rangle$ denotes the microcanonical
average calculated via the density operator
\begin{equation}
\rho = \dfrac{\delta (E-H)}{ \int \! d^{N}\mathbf{x} \,
\delta({H}(E-\mathbf{x}))} \, .
\label{rho}
\end{equation}
Therefore, the criticism of the thermodynamic consistency of the Boltzmann entropy 
raised in Ref. \cite{Dunkel2013} concerned the fact that the Gibbs
entropy ($s=s_G$) satisfies both the identities \eqref{seia} and \eqref{seib}
whereas Boltzmann entropy ($s=s_B$) does not satisfy \eqref{seib}.

\section{\label{sec:proof} Proof of consistency of the Boltzmann entropy}
In this section we  prove that the Boltzmann entropy is thermostatistically consistent.
We show that Eq. \eqref{seib} is not a fundamental condition
and it should not be satisfied in general in the microcanonical ensemble.
We also demonstrate that the Gibbs entropy is inconsistent with a different known thermostatistical
condition relating the generalized pressure and the free energy.
We finally emphasize that the entropy $s$, defined as the primitive associated with the Clausius' 
integrating factor,
coincides with $s_B$.

\subsection{ Should  the identity \texorpdfstring{Eq. \eqref{seib}}{Lg} be satisfied by the microcanonical entropy?}
In Ref. \cite{Dunkel2013} it is argued that Eq. \eqref{seib} stems from the correct identification
between thermodynamic quantities and statistical expectation values.
In particular, Eq. \eqref{seib} is derived by matching the
thermodynamical (generalized) pressure $p_\mu = - (\partial \epsilon /\partial
a_\mu)_{s,a_\nu \neq a_\mu}$ (rhs. of Eq. \eqref{seia}) to the microcanonical
average $- \langle \partial h/\partial a_\mu \rangle$ (rhs. of Eq. \eqref{seib}).
Therefore, since only the Gibbs entropy satisfies Eq. \eqref{seib}
in \cite{Dunkel2013} it is concluded that the Boltzmann entropy is inconsistent.

% It is there concluded that it is possible identify by
%consideration, in \cite{Dunkel2013} is concluded that the identity between 
%Eqs. \eqref{seia} and \eqref{seib} which prove the
%thermodynamic and mathematical consistency of Gibbs entropy and the inconsistency of the Boltzmann entropy.
%
%, since the latter does not satisfy
%such identities. Nevertheless, in our opinion the above mentioned identification
%of thermodynamic quantities with statistical expectation values, is not valid in
%the general case, as we will prove in the following part of the subsection.

We prove here that Eq. \eqref{seib} is a mathematical
property of Gibbs entropy but not a general consistency condition for the entropy in the microcanonical
ensemble.
In particular we prove that in general
\begin{equation}
\left(\dfrac{\partial \epsilon }{\partial a_\mu} \right)_{s,a_\nu \neq a_\mu}
\neq
\left\langle \dfrac{\partial h }{\partial a_\mu}\right\rangle \, .
\label{summ}
\end{equation}

Two generalized force/pressure definitions are proposed in literature: 
\begin{equation}
p_\mu = -\left( \dfrac{\partial f}{\partial a_\mu} \right)_{T,a_\nu \neq a_\mu} \, ,
\label{fmu}
\end{equation}
and 
\begin{equation}
p_\mu = - \left\langle \dfrac{\partial h}{\partial a_\mu} \right\rangle \, .
\label{Fmu}
\end{equation}
The former is derived from the thermodynamic Maxwell relations \cite{Huang_1987}, the latter is also
generally proposed in text books \cite{Landau} and it is essentially extrapolated from calculations
performed on free-particle systems confined in a box.
In the following we show that in the general case --that includes systems with negative Boltzmann
temperatures-- the correct definition is the first one Eq. \eqref{fmu}. This fact entails that the issues of
inconsistency ascribed to the Boltzmann entropy lose validity. 

Let us consider an (almost) isolated system, and let us assume that
the dynamics of any observable $O(t)$ is governed by a density of Hamiltonian $h(a(t))$ with time-dependent
external control parameters $a(t)$ through the Hamilton-Heisenberg equations
\begin{equation}
\dfrac{dO(t)}{dt} = {\cal L}[O(t),h] \, ,
\label{H-Heq}
\end{equation}
which holds for sufficiently slow parameter variations, i.e. processes that are adiabatic, and
where for classical systems the Lie-bracket ${\cal L}[O, h]$ is given by the
Poisson-bracket, whereas in the case of quantum systems the Lie-bracket corresponds to standard
commutators, ${\cal L}[O, h] = [O,h]/(i\hbar)$. 
In the case $O(t) = h(a(t))$ the Hamilton-Heisenberg equations yields
\[
\dfrac{d h}{d t} = \sum_\mu \dfrac{\partial h}{\partial a_\mu} 
\dfrac{d a_\mu}{dt} \, .
\]
By averaging over some suitably defined ensemble\footnote{For instance the time average
performed on a time interval large with respect to the fast degrees of freedom and short
with respect to the time scales of the external parameters.
Alternatively, for the microcanonical average compute according to Eq. \eqref{phimu} with the measure
of Eq. \eqref{hs-measure}, one can verify that
\begin{eqnarray}
\dfrac{d 
\left\langle h
\right\rangle
}{d t}  &=&
\left\langle
\dfrac{d h}{d t} 
\right\rangle
-
\left\langle
h \dfrac{1}{\Vert \nabla h \Vert}\dfrac{d \Vert \nabla h \Vert}{ d t}
\right\rangle
+ 
%\nonumber \\
%& &
\left\langle
h 
\right\rangle
\left\langle
\dfrac{1}{\Vert \nabla h \Vert}\dfrac{d \Vert \nabla h \Vert}{ d t}
\right\rangle = \left\langle
\dfrac{d h}{d t} 
\right\rangle \, .
\label{check}
\end{eqnarray}

}, and by identifying
$\epsilon = \langle h \rangle $, one gets \citep{Dunkel2013} SI
\begin{equation}
\dfrac{d \epsilon}{d t} = \sum_\mu \left\langle \dfrac{\partial h}{\partial a_\mu} \right\rangle 
\dfrac{d a_\mu}{dt} \, .
\label{seven}
\end{equation}
Thus, the change $d\epsilon$ in internal energy of a system whose dynamics is governed by
the Hamilton-Heisenberg equations is equal to the sum of works $\left\langle {\partial h}/{\partial a_\mu}
\right\rangle da_\mu$ 
performed on the system. 
In order to calculate the generalized pressure it is necessary to derive the total work done by
the system during the dynamically adiabatic process. In literature \citep{Dunkel2013} SI it has
been argued that a dynamically-adiabatic process, described by the Hamilton-Heisenberg equations
\eqref{H-Heq}, is an adiabatic process also in the conventional thermodynamic sense, that is
an isentropic process. Therefore, by comparing the microscopically-derived relation 
\eqref{seven} with the standard thermodynamical relations for some thermodynamic adiabatic process
($ds=0$)
\[
\dfrac{d\epsilon}{dt} = \left(\dfrac{\partial \epsilon}{\partial s} \right)_{a_\mu} \dfrac{d s}{dt} +
\sum_\mu \left(\dfrac{\partial \epsilon}{\partial a_\mu} \right)_{s} \dfrac{d a_\mu}{dt} =
\sum_\mu \left(\dfrac{\partial \epsilon}{\partial a_\mu} \right)_{s} \dfrac{d a_\mu}{dt} 
\]
one gets $({\partial\epsilon}/{\partial a_\mu} )_s = \left\langle 
{\partial h}/{\partial a_\mu} \right\rangle $
%\begin{equation}
%\left( \dfrac{\partial\epsilon}{\partial a_\mu} \right)_s =
%\left\langle \dfrac{\partial h}{\partial a_\mu} \right\rangle 
%\label{pressure}
%\end{equation}
from which the pressure definition \eqref{Fmu} comes. \\
However, only for a restricted class of systems dynamically-adiabatic processes \eqref{H-Heq} are adiabatic
also in the conventional thermodynamic sense.

In fact, in the general case, one has to consider systems with
Hamiltonians containing both a (density) kinetic term $K$ and a potential one $V$.
Thus, when along the dynamics a parameter $a_\mu(t)$ is adiabatically varied, both $K$ and $V$ vary.
During an infinitesimal time variation $dt$ we have
\[
d \epsilon= d k + d v \, ,
\]
where $k = \langle K \rangle$, $v = \langle V \rangle$ and $\langle \cdot \rangle$ is the microcanincal average.
After the kinetic energy theorem, the work $d l$ done by the system during such time is
\[
dl = - d k  \, ,
\]
thus the second thermodynamic law gives
\begin{equation}
\delta q = d \epsilon + dl = d v \neq 0 \, .
\label{stl}
\end{equation}
This fact proves that, in the general case, a process although dynamically adiabatic could be
non adiabatic ($d s \neq 0$) in the conventional thermodynamic sense, consequently Eq. 
\eqref{Fmu} has no justification.

As an example, let us consider a classical gas of harmonic oscillators. For instance, one can image of 
adiabatically varying the frequency $\omega$ of the oscillators. For this system
the equipartition theorem holds and entails
\[
k=v
\]
independently from the value of $\omega \neq 0$. Thus, for this system \eqref{stl} gives
\[
\delta q = d k \neq 0  \, .
\]

It is worth highlighting that the generalized pressure of Eq. \eqref{fmu} is derived from the Helmholtz
free energy $f$, that is the energy subtracted of the heat contribution, and, in this respect, it does
not have such issues.

Also for the class of systems with a Hamiltonian made of a kinetic term only, the Boltzmann entropy does not
have any issue of consistency.
In fact, for systems whose energy can be stored just in the kinetic term, adiabatic dynamical
processes are also adiabatic in the thermodynamical sense ($dv=0$, $d\epsilon = dk= - dl$ and
$ds=\delta q/T =0$). Nevertheless, in this case, one can prove that Eq. \eqref{fmu} reduces to Eq. \eqref{Fmu}
in the following way.
For this class of systems $({\partial\epsilon}/{\partial a_\mu} )_s = \left\langle 
{\partial h}/{\partial a_\mu} \right\rangle $ and by setting this expression in the
general pressure definition \eqref{fmu}, after having used the definition \eqref{ZTdL} with $s=const$,
one gets the expression \eqref{Fmu}.

\subsection{ Inconsistency of the Gibbs entropy}
A robust consistency condition can be derived by resorting to ensemble equivalence. In fact,
the first member of equation \eqref{seib} --in the case of a reversible transformation--
is the opposite of the generalized pressure \cite{Huang_1987}, i.e. 
\begin{equation}
p_\mu = - 
\left(\dfrac{\partial s}{\partial \epsilon} \right)_{\bf a}^{-1} 
\left(\dfrac{\partial s}{\partial a_\mu} \right)_{\epsilon,a_\nu \neq a_\mu} \, ,
\label{pmulhs}
\end{equation}
and,
\emph{from the thermodynamic Maxwell relations that are valid independently from any given
statistical ensemble} \cite{Huang_1987}, it results
\[
p_\mu = -\left( \dfrac{\partial f}{\partial a_\mu} \right)_{T,a_\nu \neq a_\mu} \, .
\]
Now, we have derived the Eq. \eqref{ZTdL} that holds when entropy is a concave function, e.g. 
for standard systems with short range interaction. Thus for this class of systems Eq. \eqref{ZTdL}
yields\footnote {Note $\epsilon^* = \epsilon^*(a_\mu)$,
$s^{^{_\infty}}_{^{_{\scriptscriptstyle B}}}(\epsilon^*) =
s^{^{_\infty}}_{^{_{\scriptscriptstyle B}}}(\epsilon^*(a_\mu),a_\mu)$.}
\begin{equation}
p_\mu = - \left( \dfrac{\partial \epsilon^*}{\partial a_\mu}   
 -\dfrac{\frac{\partial s^{^{_\infty}}_{^{_{\scriptscriptstyle B}}}(\epsilon^*)}{\partial a_\mu}}{k_B\beta} 
 -\dfrac{\frac{\partial s^{^{_\infty}}_{^{_{\scriptscriptstyle B}}}(\epsilon^*)}{\partial \epsilon^*}}{k_B\beta}
 \dfrac{\partial \epsilon^*}{\partial a_\mu}
\right) = \dfrac{1}{ k_B \beta} \dfrac{\partial
s^{^{_\infty}}_{^{_{\scriptscriptstyle B}}}(\epsilon^*)}{\partial a_\mu} \, ,
\label{good}
\end{equation} 
with the first member of Eq. \eqref{seib}, entails $s=s_B$
$ (\partial s/\partial \epsilon)_{\bf a} = k_B \beta $ and $s=s_B$.

In conclusion identity \eqref{seib} in the general case is not correct. In the case of systems of ``free''
particles in a box Eq. \eqref{seib} holds, however this is not an issue for the consistency of the
Boltzmann entropy since, in this case, Eq. \eqref{fmu} reduces to \eqref{Fmu} and,
by Eq. \eqref{pmulhs} identity \eqref{seib} results proved also for $s=s_B$.

\subsection{Clausius' integrating factor}
\label{subsec:clausius}
A further test bed for the consistency of Boltzmann entropy concerns the question of the
integrating factor for heat.
The second law of thermodynamics for the heat density $q$ reads
\[
\delta q = d \epsilon + \sum_\mu p_\mu da_\mu \, ,
\]
thus, in systems where all the statistical ensembles are equivalent, from Eq. \eqref{good} we get
\[
\delta q = d \epsilon +  \sum_\mu \dfrac{1}{ k_B \beta}\left( \dfrac{\partial 
s^{^{_\infty}}_{^{_{\scriptscriptstyle B}}}}{\partial a_\mu}
\right)_{s,a_\nu \neq a_\mu}  da_\mu
\]
which yields
\[
\dfrac{\delta q}{T} = \dfrac{d \epsilon}{T} +  \sum_\mu \left(
\dfrac{\partial s^{^{_\infty}}_{^{_{\scriptscriptstyle B}}}}{\partial a_\mu}\right)_{s,a_\nu \neq a_\mu}
da_\mu \, .
\]
Thus $1/T=\left( \partial s^\infty_{^{_{\scriptscriptstyle B}}} /
\partial \epsilon \right)$ is an integrating function for $\delta q$ according to the formulation
given by Clausius of the second law of thermodynamics.  
Consequently, the entropy $ds=\delta q /T$ defined as the primitive associated with the Clausius'
integrating factor coincides with the Boltzmann entropy. Therefore, also in this respect,
$s^\infty_{^{_{\scriptscriptstyle B}}}$ appears perfectly consistent from thermodynamic point of view.
In Ref. \citep{Campisi05} is proven that, by starting from the definition
$p_\mu = - \langle\partial_{a_\mu} h \rangle$, the microcanonically calculated differential
form $\delta q$ admits an infinite number of integrating factors that are of the form
$\partial_E g(\Omega)$, the corresponding primitive being of the form $g(\Omega)$.
In this respect, the Gibbs entropy seems to admit more solutions for the integrating factor
then the Boltzmann entropy.

While we were finalizing the present paper we became aware of the manuscript \cite{Poulter_2015} that
contains results in agreement with that of Sec. \ref{subsec:clausius}.

%
%Eq. \eqref{pressure} holds, however
%
%
%
%
%
%for which Eq. \eqref{pressure} holds, but this is not an issue for the consistency of the
%Boltzmann entropy since Eq. \eqref{fmu} reduces to \eqref{Fmu} and by Eq. \eqref{good} identity \eqref{seib}
%results proved in the case of $s_B$.

%For systems whose energy can be stored just in the kinetic term, adiabatic dynamical processes are also adiabatic
%in the thermodynamical sense since, in this case, $dv=0$ and $d\epsilon = - dl$.
%However, this does not represent an issue for the thermodynamic consistency of the Boltzmann entropy.
%In fact, in this it is $d\epsilon = \sum_\mu \langle \partial_{a_\mu} h \rangle d a_\mu$

%
%In fact, in this case the potential energy defines, by means the parameters $a_\mu$, the spatial volume occupied
%by the system $a_\mu = \alpha, \beta, \gamma$ and $vol = \alpha \cdot \beta \cdot \gamma$.
%For this system it results
%\[
%\omega(\epsilon,vol) = \sigma \epsilon^{3 N/2-1} (vol)^N \, , 
%\]
%where $N$ is the number of particles and
%\[
%\sigma = \dfrac{3 (2\pi m)^{3 N/2}}{2 (N-1)! h^3 \Gamma (3 N /2 +1 )} \, .
%\]
%Thus, for instance
%\[
%T_B \left(
%\dfrac{\partial s_B}{\partial \alpha}
%\right)_{\epsilon,\beta, \gamma}
%= \dfrac{\epsilon N}{(3N/2-1)\alpha}
%\]

\section{\label{sec:evidence} Paradigmatic evidences}
In this section we consider two different systems supporting negative-temperature states. 
The first one is a collection of $N$ undistinguishable uncoupled $1/2$ spins in a magnetic field $B$,
like the one considered in \cite{Dunkel2013} and \cite{Campisi_2015,Campisi_2016}. This is a particular case of
the class of systems discussed in the seminal work by \cite{Ramsey}.
Next, we address a tight-binding model describing $N$ classical, or quantum, particles
hopping across the sites of a lattice of length $L$, which bears   relevance
to a recent experiment where negative-temperature states have been created for motional degrees of freedom
of ultracold atoms loaded in an optical lattice \cite{Braun2013}. 
In our calculation we assume the systems as at the thermodynamic equilibrium without
considering the dynamical process necessary to realize such equilibrium.
For both models we show that the Boltzmann microcanonical ensemble produces results that are equivalent
to those obtained in the canonical (and, possibly, grand canonical) ensemble, where the inverse temperature
is just an external parameter.

The Hamiltonian for the first system is
\begin{equation}
H = - m B \sum_{j=1}^N \sigma_j 
\end{equation}
where $m$ is the magnetic moment of the individual spin and $\sigma_j = \pm 1$.
The canonical partition function for this system is easily evaluated as
\begin{equation}
Z = 2^N \cosh^N (\beta m B) = e^{-\beta f} \, ,
\end{equation}
where $f$ is the Helmoltz free energy. The internal energy and the entropy are then
\begin{align}
\label{Espin1}
E = & -\frac{\partial}{\partial \beta} \log Z = -N m B \tanh (\beta m B) \, ,\\
\label{Sspin1}
S = & k_{\rm B} \beta^2 \frac{\partial}{\partial \beta} f = k_{\rm B} N\Big[\log 2 + 
\log (\cosh (\beta m B)) \nonumber \\
 &- \beta m B \tanh(\beta m B) \Big] \, .
\end{align}
Inverting Eq.~\eqref{Espin1} for $\epsilon = E/(N m B)$, where $\epsilon \in [-1,\, 1]$,
and plugging the result into Eq.~\eqref{Sspin1} yield
\begin{align}
\label{Espin2}
\beta = & - \arctanh(\epsilon)\\%-\atanh(h)\, , \\
\label{Sspin2}
S = & \frac{k_{\rm B} N}{2} \left[ 2 \log 2 - (1+\epsilon) \log (1+\epsilon)
 - \right. \nonumber \\
&~~~~~~~~~~~~~~~~~~~\left. (1-\epsilon) \log (1-\epsilon)\right] \, .
\end{align}
Note that in the present calculation $\epsilon$ is a dimensionless quantity and, accordingly,
$\Delta$ in Eq. \eqref{entropyboltz}  is dimensionless too. Furthermore, the constant $\Delta$
satisfies the inequality $1/N \ll \Delta \ll 1$. In fact, the latter inequality guarantees that the
energy-grid step, remains much bigger than the energy levels spacing and much smaller than the energy
band, also when $B$ and $N$ are changed. In energy units this inequality corresponds to the
following $\mu B \ll \Delta \ll \mu B N$, thus in the original physical unities $\Delta$ cannot
be maintained constant when $B \to 0$. It is worth emphasizing that, by maintaining $\Delta$ constant
when $B\to 0$ gives rise to pathological behaviours of some thermodynamical quantities derived by
$s_B$.

From Eq. \eqref{Espin2} it is evident that the temperature is positive for $\epsilon<0$ and
negative for $\epsilon>0$. Note that the entropy in
Eq.~\eqref{Sspin2} is a concave function featuring a maximum at $\epsilon=0$ and, more importantly,
its derivative with respect to $\epsilon$ coincides with the function giving the temperature at a fixed
energy density, Eq.~\eqref{Espin2}. In other words, the inverse temperature and entropy obtained
in the canonical ensemble are linked by the relation that is expected to hold true in the
microcanonical ensemble. In fact, this is a specific instance of the general relation discussed
in Sec. \ref{sec:Compar}.

It is not hard to show that Eq.~\eqref{Sspin2} coincides with the microcanonical Boltzmann entropy
$S_{\rm B} = k_{\rm B} \log(\omega) $, where $\omega(\epsilon)$ is the number of microstates
corresponding to energy density $\epsilon$. It is sufficient to observe that a state at energy density
$\epsilon$ is such that $(1-\epsilon) N/2$ spins are aligned along the magnetic field, and
$N-n = (1+\epsilon)N/2$ spins are aligned against it. Therefore
\begin{equation}
\omega(\epsilon)= \binom{N}{n} = \frac{N!}{(\frac{1+\epsilon}{2} N)!(\frac{1-\epsilon}{2} N)!}
\end{equation}
and our claim is easily proven by making use of Stirling's approximation, in view of the large number of spins.

As to the Gibbs entropy, 
\begin{equation}
\label{OmegaS}
\Omega(\epsilon) = \sum_{k=0}^{(1-\epsilon) N/2} \binom{N}{k} = \int_{-1}^{\epsilon}  \, 
\tilde\omega(\epsilon^\prime)\,d\epsilon^\prime
\end{equation}
with $ \tilde\omega(\epsilon) =\frac{N}{2} \omega(\epsilon) =  e^{\frac{N}{2} g(\epsilon)}$,
with $g(\epsilon) =\frac{2}{N} \log\left(\frac{N}{2}\right) +2 \log 2 - (1+\epsilon)
\log (1+\epsilon) - (1-\epsilon) \log (1-\epsilon)$, a concave function having a maximum at $\epsilon = 0$.
Thus, repeating the general argument illustrated in Sec. \ref{sec:Compar}, $\beta_{\rm G}(\epsilon) = 
\beta_{\rm B}(\epsilon)$ for $\epsilon<0$, and $\beta_{\rm G}(\epsilon) = 0$ for $\epsilon\geq 0$.

As second case, we consider a quantum and a classical model, of an ideal gas
of noninteracting bosons hopping on a one-dimensional lattice.
In the former case the dynamics is defined by Hamiltonian
\begin{equation}
\hat{H} =- \sum^L_{j=1} \hat{a}_j \hat{a}^\dagger_{j+1} + \mathrm{h.c.} \, ,
\label{HqHopp}
\end{equation}
where $\hat{a}_j$ ($\hat{a}^\dagger_j$) is the boson annihilation (creation)
operator at site $j$ and where periodic boundary conditions
have been assumed.
In addition to the energy, the present system has a further conserved quantity, the total number of bosons
\begin{equation}
\hat{N} = \sum^L_{j=1} \hat{a}_j^\dagger \hat{a}_j \, .
\label{Nqcons}
\end{equation} 
By plugging
$\hat{a}_j =1/\sqrt{L} \sum^{L-1}_{k=0 } \exp(-i2\pi kj/L) \hat{b}_k$ into Eqs. \eqref{HqHopp} and \eqref{Nqcons}
we get
\begin{equation}
\hat{H} =- \sum^{L-1}_{k=0} \epsilon_k \hat{b}_k \hat{b}^\dagger_{k} + \mathrm{h.c.} \, , \quad
\mathrm{and} \quad
\hat{N} = \sum^{L-1}_{k=0}\hat{b}_j^\dagger \hat{b}_j
\label{HhoppNdiag}
\end{equation} 
respectively, 
where the indices $k$ run over the dual lattice sites and $\hat{b}^\dagger_{k}$
and $\hat{b}_k$ are creation and annihilation boson operators, respectively.
The energy density levels $\epsilon_{\{n_k\}}$ of system are
\begin{equation}
\epsilon_{\{n_k\}} = L^{-1} \sum^{L-1}_{k=0} \epsilon_k n_k \, ,
\label{enrlev}
\end{equation}
where $n_k$ are integer numbers of the spectrum of $\hat{b}^\dagger_k \hat{b}_k$.
The single particle energies $\epsilon_k$ for a uniform lattice result
\begin{equation}
\epsilon_k = - 2 \cos(2 \pi k/L) \, , \quad k=0,\dots,L-1\, .
\label{epsilonk}
\end{equation}
Furthermore, each energy level has also a given total number of atoms $N=\sum^{L-1}_{k=0} n_k$.
The classical model for this system is obtained
when $b_k \to z_k$ and consistently $b^\dagger_k \to z^*_k$, where $z_k =(x_k+iy_k) \in \mathbb{C}$
($k=0,\ldots,L-1$).
Also in this case, the Hamiltonian and the total number of particles
\begin{equation}
{\cal H} =\sum^{L-1}_{k=0} \epsilon_k |z_k|^2 \, , \quad {\cal N} = \sum^{L-1}_{k=0} |z_k|^2 
\label{HN}
\end{equation}
are conserved quantities.
In the following of the present section, we compare $\beta(\epsilon)$ derived in the canonical
or grand-canonical ensembles, $\beta_B(\epsilon)$ derived in the microcanonical ensemble with the
Boltzmann entropy, and $\beta_G(\epsilon)$ derived with the Gibbs entropy. Our analysis shows
clearly a great agreement between $\beta(\epsilon)$ and $\beta_B(\epsilon)$, whereas $\beta(\epsilon)$
and $\beta_G(\epsilon)$ are absolutely irreconcilable on half of the domain of $\epsilon $.

\begin{figure}[bth]
{%\includegraphics[width=.77\linewidth]{qevb} \quad
\includegraphics[width=.77\linewidth]{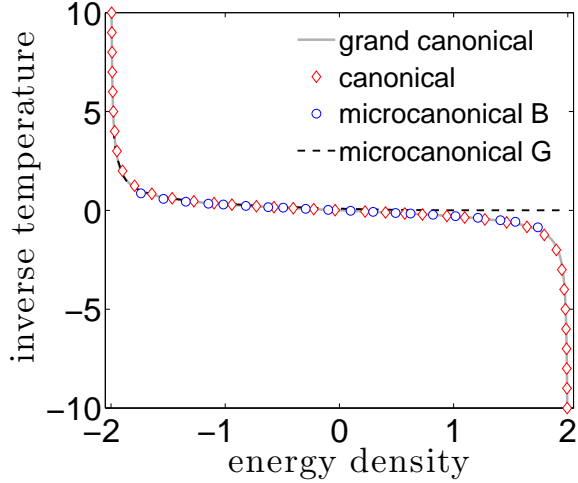}
}
\caption{
Quantum system. Relation between $\beta$ and $\epsilon$ for the three ensembles. All the curves refer
to a lattice comprising $L=20$ sites with a density of one particle per site.}
\label{qevb}
\end{figure}

\paragraph{Ideal quantum gas: grand-canonical description.}
By the canonical partition function for the quantum model
\begin{equation}
Z_N (\beta)= \sum_{\{n_k \}} 
\exp[{-\beta L \epsilon_{\{n_k\}}}] \, ,
\label{znq}
\end{equation}
where $\sum^{L-1}_{k=0} n_k = { N}$, we get
grand-partition function
\begin{equation}
{\cal Q} = \sum^\infty_{N=0} e^{\beta \mu N} Z_N(\beta) = 
\prod^{L-1}_{k=0}
\dfrac{e^{-\beta (\mu - \epsilon_k)}}
{e^{-\beta (\mu - \epsilon_k)} - 1}\, ,
\end{equation}
where the chemical potential $\mu$ has been introduced in order to fix the mean number of particles.
From the mean number of bosons in the level $\epsilon_k$ 
\begin{equation}
\langle n_k \rangle = 
\dfrac{1}{e^{\beta (\epsilon_k - \mu)} - 1}
\label{nj}
\end{equation}
we calculate the average number of bosons
\begin{equation}
N = \sum^{L-1}_{k=0} \dfrac{1}{e^{\beta (\epsilon_k-\mu)}-1} \, ,
\label{Nmubeta}
\end{equation}
and the energy density of the system
\begin{equation}
\epsilon = L^{-1} \sum^{L-1}_{k=0} \dfrac{\epsilon_k }{e^{\beta (\epsilon_k - \mu)} - 1 } \, .
\label{Emubeta}
\end{equation}
After Eq. (\ref{nj}), the condition $\langle n_k \rangle \geq 0$ imposes the constrain
$\beta (\epsilon_k-\mu) > 0$ that can be satisfied in two cases: First,
when $\mu < \epsilon_k$ ($k=0,\ldots,L-1$),  $\beta > 0$; Second, for $\mu > \epsilon_k$ ($k=0,\ldots,L-1$)
necessarily it results $\beta < 0$. Hence, in the latter case, we observe an inversion of population, namely
$\langle n_k \rangle < \langle n_k^\prime  \rangle$ with $\epsilon_k^\prime > \epsilon_k$.
For a given value of $N/L$, the inverse temperature $\beta$ is a function of the energy density $\epsilon$,
in fact, by using Eq. (\ref{Nmubeta}) and (\ref{Emubeta}) it is possible to getting rid of the chemical
potential and $\beta$ is thus expressed as a function of $\epsilon$.
Figure \ref{qevb} shows (gray) numerical results for $\beta$ vs $\epsilon$ for the case $a=1$ with $L=20$ sites
where it is evident that positive and negative values of $\beta$
are allowed.

\paragraph{Ideal quantum gas: canonical description.}
From the partition function \eqref{znq} it is possible to compute the
average of the energy density as a function
of $\beta$.
We have done this numerically by generating all the microscopic configurations with
$N=L=20$, and by averaging the density energy \eqref{enrlev} with respect to the canonical weight 
\[
\dfrac{e^{-\beta L \epsilon_{\{n_k\}}}}{Z_N(\beta)} \, ,
\]
as a function of $\beta$. The resulting curve is shown in Fig.~\ref{qevb} (red).

\paragraph{ Ideal quantum gas: microcanonical description.} 
In this case we have to calculate the density of states $\omega_N (\epsilon)$ at energy density $\epsilon$
for a system with $N$ particles.
We have obtained an approximation to $\omega_N (\epsilon)$ by binning the energies of all the
configurations with $N=L=20$, which we generated as described above.
In Fig. \ref{qevb} we compare the inverse temperature $\beta_B$ vs the energy
density $\epsilon$ (blue), obtained from the Boltzmann entropy and $\beta(\epsilon)$
derived in the grand-canonical (gray) and in the canonical ensemble (red) for the case of $N=L=20$. 
Already for this small system size it is evident
the great agreement of $\beta(\epsilon)$ between the case of Boltzmann definition and the 
corresponding relations derived in the grand-canonical and canonical ensembles.
As we have recalled earlier, the Gibbs entropy yields a non-negative inverse
temperature, irrespective of the energy density, $\beta_G(\epsilon)>0$.
Now we show that the condition $\beta >0$ and $\epsilon >0$ cannot be satisfied in the grand canonical ensemble.
Since $\langle n_k \rangle \geq 0$ for all $k$, from Eq. (\ref{nj}) we deduce $\beta (\epsilon_k -\mu) > 0$ and,
given that $\beta > 0$ necessarily $(\epsilon_k - \mu)>0$, and, in this manner $\epsilon_k > \epsilon_k^\prime$ implies
$\langle n_k^\prime \rangle >\langle n_k \rangle$. Furthermore, the single particle density levels $\epsilon_k$ have
zero average ($\sum_k \epsilon_k =0$), therefore for this weighted average we get
\[
\sum^{L-1}_{k=0} \epsilon_k \langle n_k \rangle < 0 \, ,
\] 
which proves our assertion.
As it is clearly shown in Fig. \ref{qevb}, $\beta(\epsilon)$ (gray) derived with the grand canonical ensemble and
$\beta_G(\epsilon)$ (black) derived with the Gibbs entropy are absolutely irreconcilable in the region of
$\epsilon >0$.

\paragraph{ Classical limit of the ideal quantum gas: grand canonical description.}
For the classical model the grand canonical partition 
\begin{align}
{\cal Q}_c =
%& \int \prod_{k=0}^{L-1} dx_k dy_k  e^{-\beta (\epsilon_k-\mu) (x_k^2+y_k^2)} = 
%\nonumber \\
%%\prod_{K=0}^{L-1}  \int_{-\infty}^{\infty} d p dq  e^{-\beta (\epsilon_k-\mu) (q^2+p^2)} =
%%\pi^L  \prod_{K=0}^{L-1} \int_{0}^{\infty} d n e^{-\beta (\epsilon_k-\mu) n} =
& \prod_{k=0}^{L-1} \frac{\pi}{\beta (\epsilon_k-\mu)} \, ,
\label{cgc}
\end{align}
yields
\begin{equation}
\langle n_k\rangle = \langle |z_k |^2\rangle =
%-\frac{1}{\beta}\frac{\partial}{\partial \epsilon_k} \ln {\cal Q}_c = 
%-\frac{1}{\beta}\frac{\partial}{\partial \epsilon_k} \sum_{k'=0}^{L-1} \left[\log \pi -\log \beta -
%\log(\epsilon_{k'}-\mu) \right] = 
\frac{1}{\beta} \frac{1}{\epsilon_k -\mu} \, .
\label{nk}
\end{equation}
Hence the average energy density is
\begin{equation}
\epsilon=\frac{1}{\beta L}\sum^{L-1}_{k=0}  \frac{\epsilon_k}{\epsilon_k -\mu}
\label{muE}
\end{equation}
where the chemical potential $\mu$ is determined by the condition
\begin{equation}
N = \frac{1}{\beta} \sum^{L-1}_{k=0} \frac{1}{\epsilon_k -\mu}  \, ,
\label{muNN}
\end{equation}
and it is $\mu<\epsilon_k$ ($k=0,\ldots,L-1$) in the region of positive temperatures, whereas
for negative-temperature it results $\mu>\epsilon_{k}$\footnote{Notably, Eq. (\ref{nk}) is
the classical limit ($a\gg 1$) of the quantic result in
Eq. (\ref{nj}).}.
Thus, one can derive the energy density
\begin{align}
\label{muEk}
\epsilon 
%&= \frac{1}{\beta} \sum_{k=0}^{M-1} \frac{\epsilon_k}{\epsilon_k -\mu} =
%%% \frac{1}{\beta} \sum_{k=0}^{L-1} \left[1+\frac{\mu}{\epsilon_k -\mu} \right] 
%%= \frac{M}{\beta} + N \mu 
%\frac{1}{\beta}\sum_{k=0}^{M-1} \left[ 1 + \dfrac{\mu}{\epsilon_k -\mu} \right]
=\frac{1}{\beta L} \left[ 
L + \mu N \beta \right]
\, .
\end{align}

In the thermodynamic limit $N,L\gg 1$, we can consider the continuous limit for our system
\begin{align}
\label{muN}
N &= \frac{1}{\beta} \sum_{k=0}^{L-1} \frac{1}{\epsilon_k -\mu} \approx
%%\frac{1}{\beta}  \int_{0}^{L-1} dk \frac{1}{-2\cos \left(\frac{2\pi}{L} k\right)-\mu} = \\
% \frac{1}{\beta}  \frac{M}{2\pi}\int_{0}^{2\pi } d\kappa
% \frac{1}{-2\cos \left(\kappa \right)-\mu} \nonumber\\
%&= \frac{1}{\beta}  \frac{L}{2\pi}  {\rm sign}(\beta) \frac{2 \pi}{\sqrt{\mu^2-4}} =
\frac{L}{|\beta| \sqrt{\mu^2-4}} \, .
\end{align}
Solving for the chemical potential we get
\begin{equation} 
\mu = - 2\, {\rm sign}(\beta) \sqrt{1+\frac{1}{4 \beta^2 a^2}} \, ,
\label{mufr}
\end{equation}
where $a=\frac{N}{L}$ is the particle density.
Plugging Eq. (\ref{mufr}) in the continuous limit of Eq. \eqref{muE} we get
\begin{align}
\epsilon
%\equiv \frac{E}{N} 
%&= \frac{1}{a \beta }+ \mu
= 
 \frac{1}{\beta } - 
 2 a \, {\rm sign}(\beta) \sqrt{1+\frac{1}{4 \beta^2 a^2}} \, .
\label{epsilonmu}
\end{align}
By Eqs. \eqref{muEk} and \eqref{epsilonmu} we get
\[
\mu = \dfrac{\epsilon^2 + 4 a^2}{2a\epsilon}
\] that plugged in \eqref{muEk} yields
\begin{equation}
\beta = - \dfrac{2 \epsilon}{(4 a^2 - \epsilon^2)} \, ,
\label{betaep}
\end{equation}
in which is evident that $\beta(\epsilon)$ and $\epsilon(\beta)$ are smooth functions.
Therefore, as expected, the energy density can be used to determine the
inverse temperature and chemical potential
of the system at equilibrium. 

A few comments are worthwhile. The expression for the grand canonical partition function,
Eq. (\ref{cgc}), could suggest that the point $\beta = 0$ corresponds to a singular point
where some kind of phase transition takes place. This is not the case. Indeed it is
easy to show from Eq. \eqref{mufr} that for $\beta \to 0$, $\beta \epsilon_k \to 0$ but
$\beta \mu \to -(a)^{-1}$ and, hence, ${\cal Q}_c$ does not diverge at $\beta= 0$.
\begin{figure}[bth]
{
\includegraphics[width=.77\linewidth]{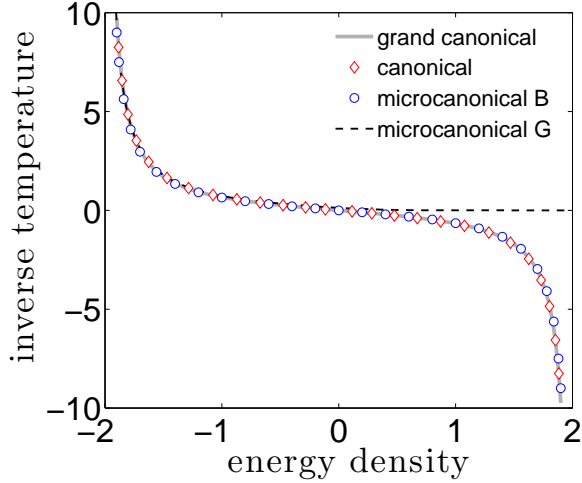}
} \quad
\caption{
Classical system.
Relation between $\beta$ and $\epsilon$ for the three ensembles. All the curves refer
to a lattice comprising $L=20$ sites with a density of one particle per site.}
\label{qcevb}
\end{figure}
In figure \ref{qcevb} it is plotted the curve \eqref{betaep} (gray).

\paragraph{ Classical limit of the ideal quantum gas: canonical description.}
The basic object in this ensemble is the canonical partition function 
\begin{equation}
Z_N(\beta) = \int d \bfx \, d \bfy \; e^{-\beta \sum_{k=0}^{L-1} \epsilon_k (x_k^2+y_k^2)}
\delta\left[N - \sum_{k=0}^{L-1}  (x_k^2+y_k^2) \right] \, ,
\label{Zcan}
\end{equation}
which, in the case of an even number of sites $L$, can be recast as
\begin{equation}
Z_N(\beta) =  \frac{\pi^{L} }{\beta^{(L-1)} L^2} B_N(\beta)
\label{ZcanI-1}
\end{equation}
where 
\begin{equation}
B_N(\beta)=-\dfrac{1}{2}
\sum^{L -1}_{k=0} \left[ \beta (4-\epsilon_k^2) N\,
+ \epsilon_k  
 \right] e^{-\beta \epsilon_k N}
 \, .
\label{ZcanI-2}
\end{equation}
The energy density in this case is
\begin{equation}
\epsilon = \dfrac{L-1}{L \beta } - \dfrac{1}{L} \dfrac{\partial}{\partial \beta} \ln B_N(\beta)
\label{eb-can}
\end{equation}
In Fig. \ref{qcevb} (red) we show the curve $\beta(\epsilon)$ derived by numerically solving
Eq. \eqref{eb-can} for $\beta$.

\paragraph{ Classical limit of the ideal quantum gas: microcanonical description.}
We want to calculate the Boltzmann entropy for the classical system introduced above.
According to Boltzmann, the entropy depends on the density of states
\begin{align}
\label{OmegaB}
&\omega(\epsilon,N) = 
\nonumber \\
&\int d \bfx \, d \bfy \; \delta\left[\epsilon N - \sum_{k=0}^{L-1} \epsilon_k (x_k^2+y_k^2) \right] \; 
\delta\left[N - \sum_{k=0}^{L-1}  (x_k^2+y_k^2) \right] 
\end{align}
through Eq. \eqref{entropyb}.
By a direct calculation, in the case of a lattice of $L=2 \ell$ sites, one gets
\begin{equation}
\omega(\epsilon,N)  =\frac{\pi^{L} N^{L-2} }{(L-2)! L^2} \ \chi(\epsilon,N)
\label{O3}
\end{equation}
where $\epsilon_0 < \epsilon < \epsilon_\ell$, the single particle energies $\epsilon_0,\epsilon_\ell$
are defined in \eqref{epsilonk}, and
\begin{align}
&\chi(\epsilon,N) =
%\Big\{
(\epsilon - 
%\right.
%\left.
 \epsilon_0)^{L-2}
%\Theta (\epsilon -\epsilon_0)
-
\nonumber \\ 
& \sum^{\ell -1}_{k=1,\epsilon_k <\epsilon} \!\!\!\!\!\!
\left[ (L-2) (\epsilon_\ell-\epsilon_k)(\epsilon_k -\epsilon_0)(\epsilon - \epsilon_k)^{L-3}\,
+ \epsilon_k (\epsilon - \epsilon_k)^{L-2} 
 \right]
%\Theta(\epsilon -\epsilon_k)
%\Big\}
\label{O3b}
\end{align}
from which it is possible to derive $\beta_B (\epsilon )$ by means the standard definition
$\beta_B = \partial_\epsilon \omega/\omega$.
Fig. \ref{qcevb} compares the inverse microcanonical temperature $\beta_B$ vs $\epsilon$ (blue) for a lattice
of $L=20$ sites and one particle per site, Eq. \eqref{betaep} obtained in the grand canonical ensemble (gray)
and the analogue relation derived in the
canonical ensemble (red).
Fig. \ref{qcevb} shows beyond a shadow of a doubt the agreement between the functions $\beta (\epsilon)$
derived from the Boltzmann's definition, and the one in the grand canonical ensemble.
In particular they both predict
negative temperatures in the domain of positive-energy densities.
Furthermore from \eqref{O3b} we have derived $\Omega(\epsilon,N)$ from which it is possible to derive
the inverse Gibbs temperature by means the standard definition $\beta_G = \omega/\Omega$.
In Fig. \ref{qcevb} we show the curve $\beta_G(\epsilon)$ (black) derived in such way. Also for the classical
model, $\beta_G(\epsilon)$ does not agree with the curves $\beta(\epsilon)$ obtained in the grand canonical
and canonical ensembles.

For the first system considered in this section, we have shown that $\beta(\epsilon)$ derived within the
canonical description agrees with $\beta_B(\epsilon)$ derived within the Boltzmann microcanonical description.
Furthermore, we have considered a second system, an ideal gas both in the classical and in the quantum case,
and we have shown that $\beta(\epsilon)$ derived within the grand canonical and
canonical descriptions agree with the same quantity derived within a microcanonical description \`a
la Boltzmann, whereas are irreconcilable with the analogues quantity derived using the Gibbs entropy.
We have shown that, for these systems the ensemble equivalence holds true provided that the Boltzmann
entropy is used within microcanonical ensemble.
Furthermore, we have seen that for the classical case of the second system, the grand canonical approach gives an explicit
form for $\beta(\epsilon)$, i.e. Eq. \eqref{betaep}.
In view of the clear agreement between the grand canonical and the microcanonical result, we can
conclude
\begin{equation}
\omega(\epsilon) \approx \omega^0 \left(4 a^2 - \epsilon^2 \right)^L \, ,
\label{omega_eps}
\end{equation}
where $\omega^0 = \exp(L s^0_B/k_B)$ does not depend on $\epsilon$.
Plugging $s_B = s^0_B + k_B \ln \left(4 a^2 - \epsilon^2 \right)$ and $\tilde{\epsilon} =0$ 
in the Eqs. \eqref{eq:Omega1} - \eqref{eq:OmegaLT2}, we deduce:
First, $T_B(\epsilon)$ is well defined within the whole range of value of
$\epsilon\neq 0$, Second, for $L\to \infty$ $T_G(\epsilon) \to T_B(\epsilon)$ in $\epsilon < 0$,
Third, in the thermodynamic limit $T_G$ is well defined only in the domain of $\epsilon$
corresponding to positive temperatures $T_B$ and it is infinity for $\epsilon \geq 0$.
This fact has dramatic consequences about the thermodynamic consistency of Gibbs entropy,
as we will show in Sec. \ref{sec:equipar}.

\emph{In the light of these facts, it is evident that thermodynamics derived from Boltzmann
entropy is perfectly consistent, both mathematically and thermodynamically, with the
thermodynamics derived in the grand-canonical and canonical ensembles, whereas the thermodynamics
derived from Gibbs entropy is inconsistent with that of these latter ensembles.}

\section{ \label{sec:equipar} Equipartition theorem}
While the Hamiltonian dynamics takes place on the phase-space hypersurface corresponding
to a given value of the energy density (and possibly of the other conserved quantities),
the Gibbs entropy requires measures involving all the energy level sets with density energy
below such value. It is therefore not immediately clear how $\beta_G$ can be measured as
an ensemble or time average. The usual answer to this question given by the supporters of
the Gibbs entropy is to use the equipartition theorem. This can be cast into the form
\begin{equation}
\beta^{-1}_G =   \langle x_k \dfrac{\partial H}{\partial x_k} \rangle
 \, ,
\label{eqthmc}
\end{equation}
where $x_k$ denotes any component of the set of dynamical coordinates and the angle brackets
denote the standard microcanonical average. Nevertheless Eq. \eqref{eqthmc} is not valid
exactly in the case of systems that admit negative temperature. For instance, in the
case of system \eqref{HN} reported in Sec. \ref{sec:evidence}, in the 
region $\epsilon > 0$ (corresponding to negative Boltzmann temperatures) the r.h.s.
of Eq. \eqref{eqthmc} with $x_k=z_k$ is $ \epsilon_k \langle |z_k|^2 \rangle$ which is
a well defined quantity for any system size $L$.
On the contrary, as we have proved above, l.h.s. goes to infinity as $L$ increases,
therefore Eq.~\eqref{eqthmc} cannot be satisfied. The reason of this failure of the
equipartition theorem, in the case of systems that admit negative temperatures,
stems from ignoring boundary terms in the derivation of Eq. \eqref{eqthmc}.
For instance, in the case of systems with bounded energy spectrum, like
the systems admitting negative Boltzmann temperatures, 
the identity of Eq. \eqref{eqthmc} is no more valid and must be corrected with
\begin{equation}
\langle x_j \dfrac{\partial H}{\partial x_k} \rangle =
\dfrac{\delta_{jk}}{\beta_G} - \dfrac{1}{\omega} \int d {\bf x} \partial_j 
\left[
x_k \Theta(E-H)
\right]
\, ,
\label{eqthmc2}
\end{equation}
that includes boundary terms. In fact, such terms in the case of systems with bounded
energy spectrum (and/or bounded coordinates), can be not null, at variance with
standard systems where $x \to \infty$ yields $H \to \infty$.

In the case of standard systems, i.e. with unbounded energy spectrum,
Eq. \eqref{eqthmc} holds, but in this case the thermodynamic quantities derived from the
Gibbs entropy differ from those obtained with the Boltzmann's definition of entropy
for quantities which are irrelevant from the statistical point of view, since they vanish
in the thermodynamic limit.

In the case of the classical model of lattice ideal gas \eqref{HN}, by a direct calculation one can show
that even in the canonical ensemble the equipartition theorem does not have the celebrated
form of Eq.~(\ref{eqthmc}), with the microcanonical average replaced by the canonical one,
but the following
\begin{equation}
\langle x_k \partial_j H \rangle_c =
%\delta_{j,k}\epsilon_k \dfrac{\epsilon_k N}{B_N} 
-\delta_{j,k} \dfrac{\epsilon_k}{\beta}
\dfrac{\partial\ln B_N (\beta) }{\partial {\epsilon_k}} \, .
%\left\lbrace
%\delta_{0,k} e^{-\beta \epsilon_0 N} -
%\delta_{\ell,k} e^{-\beta \epsilon_\ell N} -
%\sum^{\ell -1}_{j=1}
%\left[
%\beta N (4-\epsilon^2_j) + 3 \epsilon_j - \dfrac{1}{\beta N}
%\right] \delta_{j,k} e^{-\beta \epsilon_j N}
%\right\rbrace
\label{eqthcan}
\end{equation}
where, with a glance at \eqref{ZcanI-2}, the dependence on the mode index $k$ is evident.
For the same system \eqref{eqthmc2} is
\begin{equation}
\langle x_k \partial_j H \rangle = - \frac{1}{2}\delta_{j,k} \dfrac{N}{L-1}
\dfrac{\partial_{\epsilon_k} \chi(\epsilon,L)}{\chi(\epsilon,L+1)} \, .
\label{eqthmca}
\end{equation}
Therefore, for systems with bounded energy spectrum,
the equipartition theorem assumes an unexpected mathematical form which
is perfectly defined within the Boltzmann description.
On the contrary, identity \eqref{eqthmc} becomes meaningless for this class
of systems proving in such a way the flimsiness of Gibbs microcanonical thermodynamics. 
 
{\em We conclude that the identity \eqref{eqthmc} cannot be advocated
as proof in favour of Gibbs entropy, since it is not valid in the case of
systems where Gibbs and Boltzmann disagree.
The correct identity \eqref{eqthmc2}, shows that
there is not equipartition. Furthermore, since Eq. \eqref{eqthmc} is not valid, it cannot
used to measure the Gibbs temperature as a microcanonical average}.

\section{ \label{sec:measuring} Measuring temperature}
Making a temperature measurement on a system brings about, inevitably, a contact between the
system and a second ``small system''.
Especially in the present context a particular care has to be employed when we choose a second
calibrated system (thermometer) to attach to the first one (sample) in order to determine the
temperature of the latter. The thermometer has to detect the sample temperature without
destroying its equilibrium. 
This means that a thermometer capable of sustaining negative temperatures must be employed
with a sample admitting negative temperatures. Indeed, the whole system,
obtained by glueing together  a ``small'' system with unbounded energies, like an harmonic oscillator,
to the sample would be a system with unbounded energies, that is without negative temperatures.
In other words, a ``small'' system with unbounded energies will be able to detect only the positive
temperatures.
This aspect that appears as a good reasoning, has been diffusely discussed by Ramsey in its paper
\cite{Ramsey} even if several authors \cite{Dunkel2013, Romero-Rochin} seem to
have missed this point.

Furthermore, it makes sense to ask oneself if (and how) two systems admitting
negative temperatures reach equilibrium when joined together at an initial different temperature.

\begin{figure}[bth]
{
\includegraphics[width=.77\linewidth]{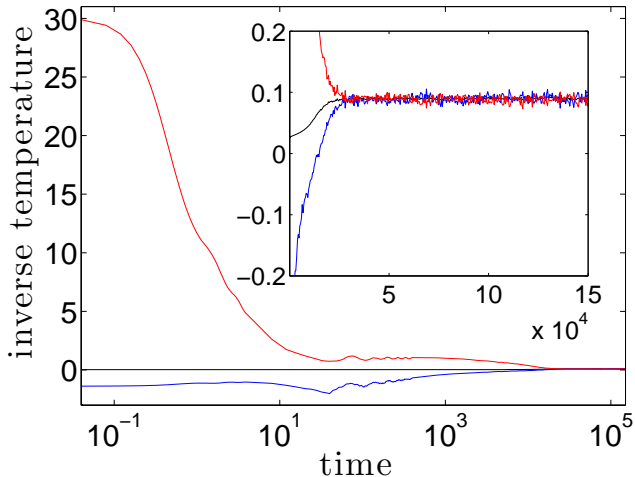}
} \quad
\caption{
The figure shows the thermalization of the subsystems $1$ and $2$ after  that they are attached at initially
different inverse temperatures. The black line is the evolution of the inverse temperature of the whole
system, the blue line is the inverse temperature of system $1$ and the red line is the inverse temperature
of the subsystem $2$. The inset shows a zoom of the last part of evolution.}
\label{thermalization}
\end{figure}

In particular, a relevant question is whether the two joined systems reach eventually the same inverse
temperature and if, or not, this latter is intermediate respect to the initial inverse temperatures
of the two systems as we expect from statistical mechanics of positive temperatures
\cite{Huang_1987}.
In order to directly verify if the Boltzmann temperature complies with this requisite, we have
simulated an experiment in which two different systems that admit negative temperatures, at different initial
temperatures, are brought to contact with each other.
Thus we have considered two systems described by the following Hamiltonians
\begin{equation}
H_{j} = - \sum_{{\bf r} {\bf r}^\prime} z^*_{\bf r} A_{{\bf r}{\bf r}^\prime} z_{{\bf r}^\prime}
-{ U}_{j} \sum_{\bf r} \log (1+|z_{\bf r}|^2)
 \, , \ {j=1,2} \, ,
 \label{DNSE}
\end{equation}
where, for each system, the indices ${\bf r}$ and ${\bf r}^\prime$ run over a two-dimensional lattice
and $A_{{\bf r}{\bf r}^\prime}$ is the adjacency matrix that describes the nearest-neighbour interaction
in two spatial dimensions.
In \eqref{DNSE} we have added to a kinetic a term similar that of Hamiltonian \eqref{HN},
an onsite nonlinear potential, in order to make the systems not integrable. 
Note, that also with the addition of this latter term, the system admits negative temperatures.
We started with the two systems
isolated with each other. In our simulations we have prepared the initial configuration of equilibrium for
the two systems at different temperatures ($\beta^0_{j}$, $j=1,2$), by means of a long time integration of
the equations of motion of the two separated systems. The inverse temperatures $\beta^0_{j}$ ($j=1,2$),
have been measured with Eq. \eqref{betafranz}, that descends from the Boltzmann entropy.
The two systems of $64 \times 64$ sites have been joined to form a
single lattice. In the simulation reported in Fig. \ref{thermalization} we set
$U_1 = 0.1$ and $\beta^0_1 \approx -1.38$, $U_2 = 0.75$ and $\beta^0_2 \approx 29.89$.  
We integrated the equations of motion of the whole system.
In Fig. \ref{thermalization} we report the inverse temperature for the whole system (black), subsystem $1$
(blue) and subsystem $2$ (red).  As it was expected, we observe that the inverse
temperature of the whole system, after a short transient, reaches an asymptotic value $\beta^f \approx 0.09$
intermediate between the initial values of the inverse temperatures of the two original systems.
Also, the inverse temperatures of the two subsystems approach the value of $\beta(t)$ along the time.
Furthermore, this value remains stable on long time scales.
For detail about these numerical results we refer to \cite{Buonsante_2015}, where we have presente
analytical and numerical evidence that Boltzmann microcanonical entropy
allows the description of phase transitions occurring at (negative Boltzmann temperatures)
high energy densities, at variance with Gibbs temperature.

It is worth remarking that, whereas this process of thermalization is well explained with the Boltzmann
temperature, we cannot say the same for the inverse temperature of Gibbs for which it is $\beta^0_1  =
\infty$ with $\beta^0_2 \approx 29.89$ and $\beta^f < \infty$.

Finally we have verified that, irrespective of the sign of
the temperature, a large lattice (that realizes a microcanonical ensemble) acts as a thermostat for
a small sublattice (that realizes a grand canonical ensemble) and that the temperatures measures for the
two systems agree, thus confirming the equivalence between the the microcanonical and the
grand canonical ensemble.

\section{ Final remarks}
We have addressed the question of the right definition of microcanonical entropy.

For systems for which the equivalence of the statistical ensembles is verified we have shown that the correct
map between the canonical average of the energy (also for systems with one or more conserved quantities)
and the Lagrangian parameter $\beta$ is that descending from the Boltzmann entropy.
Moreover, we have concluded that the only consistent definition for the microcanonical entropy is that of Boltzmann.
In fact, while for standard systems both these entropies lead to equivalent thermodynamic results in the
thermodynamic limit \cite{TodaKuboSaito}, in the case of systems with bounded energy spectrum, negative
Boltzmann temperatures are admitted, and the two microcanonical entropies lead to irreconciliable results.
In particular, when the latter circumstance is verified, the inverse temperature derived by the Gibbs
entropy coincides with the one of Boltzmann within the region of energy density where the latter is
positive, and is identically null where the Boltzmann temperature is negative. In this way, it could
happen that in correspondence of the energies where $\beta_B$ changes sign, $\beta_G$ is not a
differentiable function of $\epsilon$. But this conflicts with
the fact that the canonical and grand canonical  partition functions are smooth functions of $\epsilon$
in correspondence of such points. On the contrary, $\beta_B(\epsilon)$ is a smooth function of $\epsilon$,
and no consistence issue of this kind arises for Boltzmann entropy. 

For a general system, we have proved that: i) the Boltzmann entropy is
thermostatistically consistent;
ii) the Eq. \eqref{seib}, that has been adduced as thermostatistical self-consistency condition
for entropy \cite{Dunkel2013}, actually is not a fundamental condition for the microcanonical entropy;
iii) the Gibbs entropy is inconsistent with the thermostatistical
condition that relates the generalized pressure and the free energy.

For all these reasons we conclude that the
correct definition for the microcanonical entropy is the one of Boltzmann. 

%\appendix*

\section*{Acknowledgments}
We are grateful to J. Dunkel, S. Hilbert, P. H\"anggi and M. Campisi for 
sharing with us their different view on the Boltzmann and Gibbs entropy.
We thank M. Gabbrielli for useful discussions.

%%%%%%%%%%%%%%%%%%%%%%%%%%%%%%%%%%%%%%%%%%%%%%%%%%%%%%%%%%%%%%%%%%%%%%
% BIBLIOGRAPHY
%\begin{thebibliography}{99}
%
%%%%%%%%%%%%%%%%%%%%% Discrete and Gap solitons %%%%%%%%%%%%%

\end{document}